\documentclass[a4paper, 11pt]{article}
\pdfoutput=1
\usepackage[pdftex,
            pdfauthor={Thomas Gorochowski, Mario di Bernardo, Claire Grierson},
            pdftitle={NetEvo: A computational framework for the evolution of dynamical complex networks},
            pdfkeywords={dynamical networks; evolution; computational framework},
            pdfsubject={Network Evolution},
            pdfcreator={LaTeX, TextMate and Mac OS X}]{hyperref}

\usepackage{amsmath}
\usepackage{amssymb}
\usepackage{booktabs}
\usepackage{graphicx}
\usepackage{subfig}
\usepackage{authblk}
\usepackage{abstract}
\usepackage{caption}
\usepackage{palatino}
\usepackage{listings}
\usepackage{color}
\usepackage{hyperref}

\begin{document}
\renewcommand\Affilfont{\itshape\fontsize{9}{11}\selectfont}
\title{NetEvo: A computational framework for the evolution of dynamical complex networks}
\author[1]{Thomas E. Gorochowski}
\author[2,3]{Mario di Bernardo}
\author[4]{Claire S. Grierson}
\affil[1]{Bristol Centre for Complexity Sciences, Department of Engineering Mathematics, University of Bristol, Bristol, BS8 1TR, UK}
\affil[2]{Department of Engineering Mathematics, University of Bristol, Bristol, BS8 1TR, UK}
\affil[3]{Department of Systems and Computer Science, University of Naples Federico II, Via Claudio 21, 80125, Napoli, Italy}
\affil[4]{School of Biological Sciences, University of Bristol, Bristol BS8 1UG, UK}
\date{\small\today}
\maketitle

\begin{abstract}
NetEvo is a computational framework designed to help understand the evolution of dynamical complex networks. It provides flexible tools for the simulation of dynamical processes on networks and methods for the evolution of underlying topological structures. The concept of a supervisor is used to bring together both these aspects in a coherent way. It is the job of the supervisor to rewire the network topology and alter model parameters such that a user specified performance measure is minimised. This performance measure can make use of current topological information and simulated dynamical output from the system. Such an abstraction provides a suitable basis in which to study many outstanding questions related to complex system design and evolution.
~
\end{abstract}

\section{Introduction}
This report does not present any scientific results but instead gives an overview of a new computational framework called NetEvo. It is the aim of this tool to provide a coherent approach for the study of complex systems that can be described using dynamical networks and which change over time due to some evolutionary process. This document introduces the functionality of the framework, gives an overview of the technical architecture, presents an example of its use and describes future directions for the work. Due to the framework being under continual development we do not discuss specific implementation details that are likely to change. Instead, we direct you to the project website \url{http://www.netevo.org} where additional information can be found.

\subsection{Motivation}
Network science has seen much interest over recent years with it providing many tools for the analysis of complex systems. During this time several characteristics have been shown to be shared by systems across domains, including scale-free degree distributions \cite{Barabasi:1999p1716} and small-world type properties \cite{Watts:1998p307}. More recently, an increased focus has been placed on dynamical processes that take place on such topologies; with a key example being the modelling of epidemics such as SARS and Swine flu \cite{Colizza:2007p1968}. Furthermore, generative models have been developed to understand the driving forces that underlie network creation \cite{Barabasi:1999p1716}. 

In contrast, relatively little effort has been made in attempting to understanding how \emph{structural}, \emph{dynamical} and \emph{evolutionary} features are pieced together within a complex system. NetEvo has been developed to help meet this need. It aims to provide a computational foundation on which to investigate how these various aspects are linked and ultimately provide a more complete view of complex systems in general. 

\subsection{Development}
The NetEvo framework is fully open-source software released under the Open Source Initiative (OSI) approved MIT license. This gives end-users the greatest flexibility in using and extending the code for their own requirements, important in research based projects. To allow for collaborative development all source code, documentation and management information is hosted by Sourceforge -- \url{http://sourceforge.net}. This is an online facility used by many open-source projects to allow for developers to productively work together, handling day-to-day management of underlying infrastructure. We see NetEvo as a collaborative effort and welcome any additional help or support from others in the complex systems community.

\section{Functionality}
NetEvo performs two main functions; \emph{simulation} and \emph{evolution} of dynamical networks. In order for this to be tailored towards a specific problem it is necessary for an end-user to provide:
\begin{enumerate}
	\item A set of \emph{component dynamics} for nodes and edges,
	\item An \emph{initial network topology} for the system,
	\item The \emph{evolutionary process} that searches for improved configurations,
	\item A \emph{performance measure} $Q$ to be used as a guide during evolution. 
\end{enumerate}
Several standard types have been included for each of these categories, such as R\"{o}sller oscillator dynamics, a rewiring evolutionary process and a synchronisability performance measure (eigenratio).

\begin{figure}
	\centering
	\includegraphics[width=8.8cm]{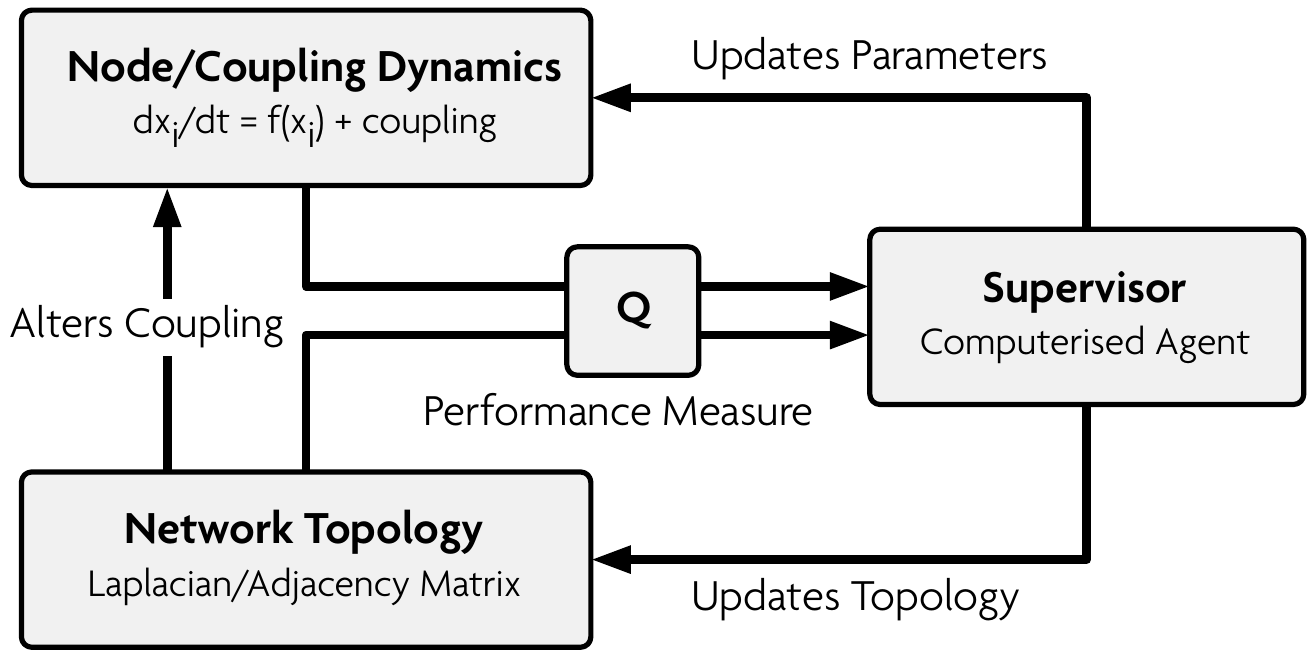}
	\caption{\label{fig:SupervisedNetwork}Flow diagram of a supervised network. The dynamical network is represented by the node/edge dynamics and network topology, while the evolutionary process is embodied in the supervisor. $Q$ represents a performance measure that the evolving network attempts to minimise.}
\end{figure}

To bring together simulation and evolution in a coherent way, the framework uses the idea of a \emph{supervisor}, illustrated in Fig.~\ref{fig:SupervisedNetwork}. Simulation is carried out by taking the component dynamics and network topology, and then numerically solving the system for a specified period of time. The solvers are a central part of NetEvo and are initially focused on systems that can be described using ordinary differential equations (ODEs).

Evolution of the system is performed by the supervisor which can be viewed as a form of optimiser. This takes as input an initial topology, simulated output from the system and user defined constraints, and aims to return an optimal or enhanced topology and parameter set. Changes to the system are assessed by using the performance measure $Q$, with smaller values representing an improved performance.

With many possible applications containing differing dynamics and constraints, it is impossible to develop a single supervisor that will work well in all situations. Instead, by default we provide a supervisor that uses a \emph{simulated annealing} meta-heruistic to search for near optimal configurations. This method has been shown to perform well for a wide range of problems with an unknown prior structure. We allow for many aspects to be customised including cooling schedule, accepting probability and halting criteria. Users are not, however, limited to only using this method. As with all of NetEvo if an alternative evolutionary process is required, possibly due to the way some physical process works, it is possible to define custom supervisors that make use of any other build-in modules. The ability to piece together various modules for a specific need is important in allowing NetEvo to be widely used.

In addition to these core features, the framework also provides a standardised file format for the description of dynamical networks. This is based on GraphML and uses attributes to hold dynamical information about various components making up the graph. With GraphML being a widely used format to exchange graph data, output from NetEvo can immediately be used by a wealth of other tools.

\section{Architecture}
The design of the framework has revolved around two main aims: \emph{efficiency} and \emph{simplicity}. These were important because firstly the computational effort required when simulating and evolving networks is large, and if we wish to investigate real-world systems in a feasible time the implementation needs to be fast. Secondly, to reduce the learning curve for end-user, developing simple well documented interfaces is vital.

\subsection{Internal Design}
\begin{figure}
	\centering
	\includegraphics[width=16.0cm]{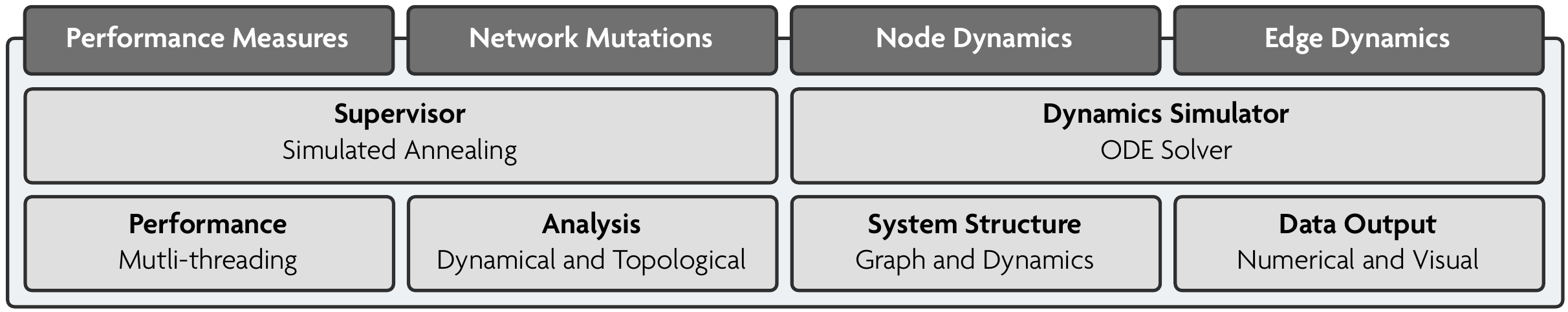}
	\caption{\label{fig:NetEvoModules}Overview of the main NetEvo modules. Core layer functionality is represented by the lightly shaded modules, while the interface layer is highlighted by the darker modules.}
\end{figure}

NetEvo has been developed solely in the C programming language due to its high performance, extensive collection of libraries, direct access key technologies (MPI, OpenMP and OpenCL) and portability across all main operating systems and computer architectures. To allow for the logical separation of components and the possible replacement of functionality based on user requirements, a modular and layered architecture has been adopted (see Fig.~\ref{fig:NetEvoModules}). 

A set of \emph{core} modules implement the fundamental features of the framework, managing the underlying data types. On top of these, \emph{interface} modules act as an intermediary to end-users, exposing the various features they can directly access. Furthermore, each individual module is compartmentalised such that alternative implementations can be provided where appropriate. This becomes important when development of a program takes place on one type of architecture, and execution on another, e.g. development on a single processor desktop computer and execution on a multi-processor computing cluster. Under these scenarios, the framework can be used with optimised modules for the respective systems and the user need not worry about the final running environment during development.

\subsection{External Dependancies}
In order to make best use of existing code and maximise the available functionality when using NetEvo, two external libraries have been directly incorporated into the core of the framework: \emph{igraph} and the \emph{GNU Scientific Library} (GSL). These were chosen due to the large functionality they provide, their code maturity and for being highly portable across computing architectures.

\subsubsection{igraph}
The igraph toolkit \cite{Csardi:2006p275} provides efficient data types with which to implement graph based algorithms for use on large data sets. The toolkit initially focused on performance, however, more recently features have diversified into areas such as community detection, visualisation and spectral analysis. Due to computational efficiency being an important factor during the design of NetEvo, igraph was selected as the internal data type for network topologies. If alternative formats are required, a wide range of common export functions are available to allow for external analysis.

\subsubsection{GNU Scientific Library (GSL)}
To provide access to a large number of numerical methods including the solving of ordinary differential equations, spectral analysis and linear algebra, the GNU Scientific Library (GSL) \cite{Galassi:2002p119} is used. Due to this library requiring GSL based data types for the input to many functions, methods have been included within NetEvo to perform necessary conversion from structural and dynamical output. 

\section{An Example}
To help illustrate the framework in action we will now present a simple example of enhancing the synchronisability of a dynamical network. To allow for our results to be comparable to those in the literature, we consider nodes having R\"{o}ssler dynamics, diffuse edge coupling and impose a fixed number of nodes equal to $100$ with a fixed degree of $4$. Evolution of the system is constrained to only take place through rewiring of existing edges and as a performance measure we chose the eigenratio which when minimised has been shown to improve the synchronisability of a system\cite{Pecora:1998p61}. The eigenratio is given by,
\begin{equation}
	Q_{ER} = \frac{\lambda_N}{\lambda_2},
\end{equation}
where $\lambda_N$ and $\lambda_2$ are the $N$th and $2$nd eigenvalues of the network Laplacian respectively. Optimising networks using the eigenratio has already been carried out in the literature \cite{Donetti:2005p229}, with resultant networks exhibiting a reduced diameter and clustering, and increased girth. By using NetEvo with this same measure we aim to show that similar topological features emerge.

In summary, we configured NetEvo in the following way:
\begin{enumerate}
	\item \emph{Component dynamics} -- R\"{o}ssler oscillator for nodes and diffusive edge coupling,
	\item \emph{Initial network topology} -- Lattice (100 nodes and 200 undirected edges),
	\item \emph{Evolutionary process} -- Edge rewiring using built-in simulated annealing supervisor,
	\item \emph{Performance measure} -- Eigenratio. 
\end{enumerate}

\subsection{Coding NetEvo}
Using the system description from the previous section we can now generate the code required to perform the network evolution. As with all programs written using NetEvo this breaks into five main steps:
\begin{enumerate}
	\item Initialisation of the framework
	
	\item Creation of the NetEvo system (node/edge dynamics and initial topology)
	
	\item Configuration of simulation and supervisor
	
	\item Evolution of network through execution of the supervisor
	
	\item Clean-up
\end{enumerate} 
The following code listing implements these steps and have been commented to help explain the exact tasks being performed.

\vspace{0.2cm}
\lstset{basicstyle=\tt\footnotesize, 
        numbers=left, 
        tabsize=2, 
        numberstyle=\tiny, 
        commentstyle=\color{blue}, 
        morecomment=[l]{/*}}
\begin{lstlisting}
/* NetEvo and igraph libraries will be used */
#include <netevo/netevo.h>
#include <igraph.h>
	
int main (int argc, const char * argv[]) {	
	
	/* Define required variables */
	FILE *gFile;
	igraph_t G;
	ne_dyn_t *D = NULL;
	ne_system_t *S = NULL;
	ne_sup_sa_params_t *saParams;
	ne_mut_t *mutFn;
	ne_ode_solver_config_t *solverConfig;
	double mParams[NE_MAX_MUT_PARAMS];

	/* Initialise the framework seeding the random number generator */
	ne_common_init(time(NULL));
	/* Configure the standard output and error logging */
	ne_output_init("EXAMPLE", ".txt", NE_LOG_FULL, FALSE, NULL, NULL);
	
	/* Read in the initial network topology from file */
	gFile = fopen("Lattice.txt", "r");
	igraph_read_graph_gml(&G, gFile);
	igraph_to_undirected(&G, IGRAPH_TO_UNDIRECTED_COLLAPSE);
	
	/* Select the node and edge dynamics and set global coupling to 0.5 */
	D = ne_dyn_alloc(&G, "Rossler3", "Diffuse3XZOnly", FALSE, FALSE, FALSE, FALSE, 3);
	D->defEdgeDyn->defParams[0] = 0.5;

	/* Create the NetEvo system */
	S = ne_system_alloc(&G, D);

	/* Define the mutation function (rewiring) */
	ne_pop_array_zero(mParams, NE_MAX_MUT_PARAMS);
	mParams[0] = 1.0;
	mutFn = ne_mut_rewire_alloc(mParams);

	/* Set parameters for the ODE solver */
	solverConfig = (ne_ode_solver_config_t *)malloc(sizeof(ne_ode_solver_config_t));
	solverConfig->length = 100;
	solverConfig->eps_abs = 10e-5;
	solverConfig->eps_rel = 10e-5;
	solverConfig->initStep = 0.0001;
	solverConfig->minStep = 0;
	solverConfig->fixedStep = FALSE;
	solverConfig->step = gsl_odeiv_step_rkf45;

	/* Set parameters for the simulated annealing supervisor */
	saParams = (ne_sup_sa_params_t *)malloc(sizeof(ne_sup_sa_params_t));
	saParams->initialTrials = (int)igraph_vcount(S->graph);
	saParams->tempReduce = 0.9;
	saParams->mainTrials = 5000;
	saParams->acceptTrials = 500;
	saParams->acceptRunsNoChange = 5;
	saParams->minTemp = 0.0000001;
	saParams->maxIterations = 500000;
	saParams->initialTempFn = &ne_sup_sa_int_temp_basic;
	/* Use the eigenratio performance measure */
	saParams->QFn = ne_per_top_eigenratio_params_alloc(NULL);
	/* Use the rewiring mutation function (defined previously) */
	saParams->mutationFn = mutFn;
	/* Use the previously defined ODE solver configuration */
	saParams->solverConfig = solverConfig;
	/* Use the standard network measures analysis function */
	saParams->analysisFn = &ne_analysis_std_measures;
	saParams->initCond = NULL;

	/* Run the simulated annealing supervisor to evolve the network */
	ne_sup_sa_run (S, saParams);
	
	/* Release any resources used by NetEvo */
	ne_output_finalise();
	ne_common_finalise();

	return 0;
}
\end{lstlisting}

\subsection{Results}
\begin{figure}[b]
	\begin{center}
	\subfloat[Diameter]{
	\includegraphics[width=3.8cm]{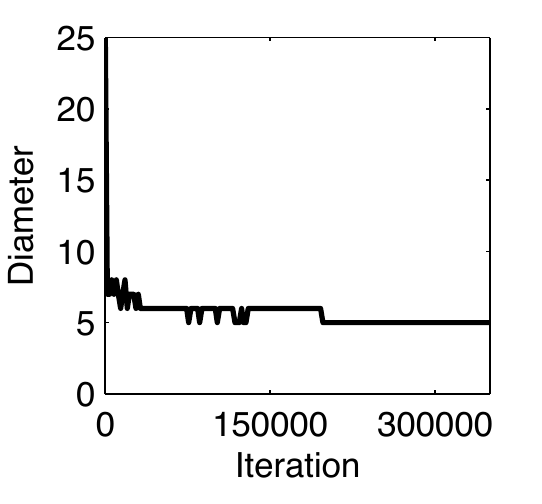}}
	\subfloat[Clustering]{
	\includegraphics[width=3.8cm]{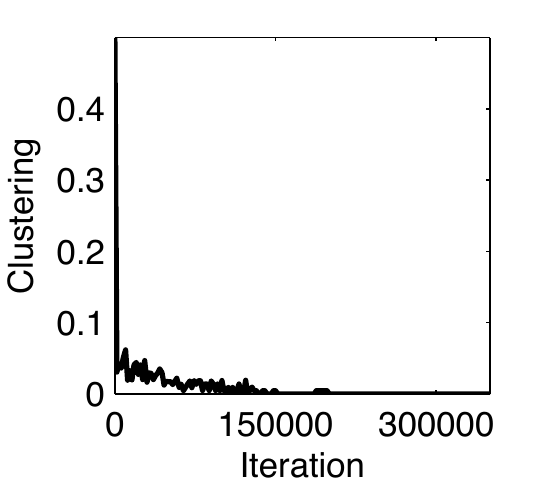}}
	\subfloat[Girth]{
	\includegraphics[width=3.8cm]{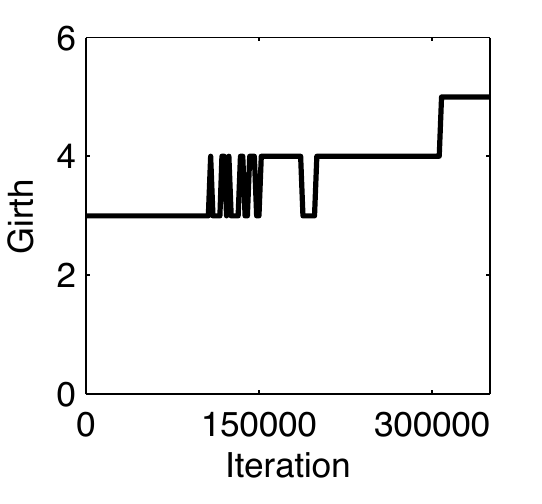}}\hspace{0.1cm}
	\subfloat[Final Topology]{
	\includegraphics[width=3.6cm]{./Topology.pdf}}
	\caption{\label{fig:Results}Results from using NetEvo to evolve networks with improved synchronisability. Network visualisation was performed using GraphViz \cite{ellson2002graphviz} and an energy minimised layout.}
	\end{center}
\end{figure}

A final topology and some standard network measures taken throughout the evolutionary process are shown in Figure \ref{fig:Results}. These illustrate similar features to those seen in the literature with decreasing diameter and clustering, and increasing girth. Visualisation of the topology also shows similar characteristics to examples presented in \cite{Donetti:2005p229}, with an \emph{entangled} structure that makes any localised structure very difficult to see.

In summary, although we have not presented any new results we have shown that NetEvo allows for evolutionary simulations to be created with very little effort and that the framework provides a flexible tool to study a wide range of problems.

\section{Future Directions}\label{sec:Future}
At the time of writing, NetEvo focuses on complex networks that can be described using continuous ordinary differential equations that are evolved using a standard simulated annealing meta-heuristic. Although a large number of problems can be expressed in this form, it does limit possible applications of the framework. The majority of future directions focus on reducing this limitation.

\subsection{Range of Dynamics} 
A set of dynamical components are provided with NetEvo to allow for many general types of system to be described, e.g. gene regulatory networks (GRNs), however these are all based on ODE representations. Some systems are not continuous or deterministic which limits the use of NetEvo. A possible extension would be to provide the ability to have discrete dynamics and simulations that include stochastic effects. This could be performed using a stochastic master equation (Gillespie algorithm) or stochastic differential equations. In addition, the ability to add delays would also be useful where interactions are not instantaneous, e.g. neural signalling.

\subsection{Enhanced Supervisors}
At present a built-in simulated annealing based supervisor is provided as a flexible all purpose method. In some cases, however, the problems being studied may have some intrinsic structure which can be incorporated into the evolutionary processes. An interesting future direction would be to develop a general purpose supervisor that attempts to \emph{learn} such structure and utilise it to favourably direct evolution of the system. This would likely involve the use of statistical machine learning, giving an additional benefit of providing a way of extracting hidden structure from systems that are currently not well understood.

\subsection{High Performance Computing}
Speed is vital to allow for NetEvo to study problems in reasonable amounts of time. Multi-threading has been incorporated during the simulation process, which allows for better use of resources on multi-core systems, however, will not provide a benefit when running on high performance computing clusters. The benefit of multi-threading is limited to the local computing node.

One possible extension would be use the Message Passing Interface (MPI) to distribute processing among computers in a cluster. Due to the overhead imposed by communication, such an approach would not be suitable for the simulation of dynamics, however, would be appropriate for population based supervisors (e.g. a genetic algorithm). In this case each individual in the population could reside on a separate compute node and be evolved in isolation during each iteration.

Alternatively, if simulation speed is important an attempt could be made to incorporate heterogeneous computing. This relates to using processors found on modern graphic cards to perform highly parallel computation and may be suitable for improving the performance of numerical simulation.
	
\subsection{Visualisation of Network Evolution}
Being able to visualise and make sense of the evolutionary process is vital in to make use of output from NetEvo. Many algorithms exist to produce aesthetically pleasing visualisations of static graphs. However, these do not always extend well to situations where the topology is changing over time. Some force based methods have been proposed \cite{Veldhuizen:2007p1589}, however, suffer stability issues for very large or highly connected graphs. Furthermore, with evolution of graphs often taking millions of iterations, any method must allow for quick interactive movement to points of interest without the need to view all intermediate steps.

\section*{Acknowledgements}
TEG acknowledges the support of EPSRC grant EP/5011214.

\small
\bibliographystyle{abbrv}
\bibliography{NetEvoTechnicalReport}

\begin{thebibliography}{1}

\bibitem{Barabasi:1999p1716}
A.~Barabasi and R.~Albert.
\newblock Emergence of scaling in random networks.
\newblock {\em Science}, 286:509--512, 1999.

\bibitem{Watts:1998p307}
D.~Watts and S.~Strogatz.
\newblock Collective dynamics of small-world networks.
\newblock {\em Nature}, 393:440--442, 1998.

\bibitem{Colizza:2007p1968}
V.~Colizza, A.~Barrat, M.~Barth{\'e}lemy, and A.~Vespignani.
\newblock Predictability and epidemic pathways in global outbreaks of
  infectious diseases: the sars case study.
\newblock {\em BMC medicine}, 5(1):34, 2007.

\bibitem{Csardi:2006p275}
G.~Cs{\'a}rdi and T.~Nepusz.
\newblock The igraph software package for complex network research.
\newblock {\em InterJournal Complex Systems}, 1695, 2006.

\bibitem{Galassi:2002p119}
M.~Galassi, J.~Davies, J.~Theiler, B.~Gough, G.~Jungman, M.~Booth, and
  F.~Rossi.
\newblock {\em GNU scientific library}.
\newblock 2002.

\bibitem{Pecora:1998p61}
L.~Pecora and T.~Carroll.
\newblock Master stability functions for synchronized coupled systems.
\newblock {\em Physical Review Letters}, 80(10):2109--2112, 1998.

\bibitem{Donetti:2005p229}
L.~Donetti, P.~Hurtado, and M.~Mu{\~n}oz.
\newblock Entangled networks, synchronization, and optimal network topology.
\newblock {\em Physical Review Letters}, 95(18):188701, 2005.

\bibitem{ellson2002graphviz}
J.~Ellson, E.~Gansner, L.~Koutsofios, S.~North, and G.~Woodhull.
\newblock {Graphviz-open source graph drawing tools}.
\newblock {\em Lecture Notes in Computer Science}, pages 483--484, 2002.

\bibitem{Veldhuizen:2007p1589}
T.~L. Veldhuizen.
\newblock Dynamic multilevel graph visualization.
\newblock {\em arXiv}, cs.GR, 2007.

\end{thebibliography}

\end{document}